\documentstyle[aas2pp4]{article}
\input psfig.tex 
\newcommand{\bc}{\begin{center}}
\newcommand{\ec}{\end{center}}

\def\@p@@sangle#1{
		\@angletrue
		\edef\@p@sangle{#1} %\number\dimen100}
}
\def\etal{et al. }

\def\ms{$M_\odot$ }
\def\msp{$M_\odot$}
\def\ls{$L_{V\odot}$ }

\def\ltsima{$\; \buildrel < \over \sim \;$}
\def\ltsim{\lower.5ex\hbox{\ltsima}}
\def\gtsima{$\; \buildrel > \over \sim \;$}
\def\gtsim{\lower.5ex\hbox{\gtsima}}

\def\frac#1#2{{#1\over#2}}           
\begin{document}

\title{A NEW APPROACH TO DETERMINE THE INITIAL MASS FUNCTION 
IN THE SOLAR NEIGHBORHOOD 
}

\author{Takuji Tsujimoto$^{1}$, Yuzuru Yoshii$^{2,4}$, 
          Ken'ichi Nomoto$^{3,4}$, Francesca Matteucci$^{2,5}$}
\vspace{-0.5cm}
\author{and}
\vspace{-0.5cm}
\author{Friedrich-Karl Thielemann$^6$, Masaaki Hashimoto$^7$}
\vspace{0.5cm}
\affil{$^1$National Astronomical Observatory,
         Mitaka, Tokyo 181, Japan}
\affil{$^2$Institute of Astronomy, Faculty of Science,
         University of Tokyo, Mitaka, Tokyo 181, Japan}
\affil{$^3$Department of Astronomy, School of Science,
         University of Tokyo, Bunkyo-ku, Tokyo 113, Japan}
\affil{$^4$Research Center for the Early Universe, School
         of Science,
         University of Tokyo, Bunkyo-ku, Tokyo 113, Japan}
\affil{$^5$Dipartimento di Astronomia, Universita di Trieste,
         SISSA, Via Beirut 2-4, I-34013, Italy}
\affil{$^6$Institut f\"ur theoretische Physik, 
         Universit\"at Basel, Basel, Switzerland}
\affil{$^7$Department of Physics, Faculty of Science, 
         Kyushu University, Fukuoka 810, Japan}

\begin{abstract}

Oxygen to iron abundance ratios of metal-poor stars provide
information on nucleosynthesis yields from massive stars which end in
Type II supernova explosions.  Using a standard model of chemical
evolution of the Galaxy we have reproduced the solar neighborhood
abundance data and estimated the oxygen and iron yields of genuine SN
II origin.  The estimated yields are compared with the theoretical
yields to derive the relation between the lower and upper mass limits
in each generation of stars and the IMF slope.  Independently of this
relation, we furthermore derive the relation between the lower mass
limit and the IMF slope from the stellar mass to light ratio in the
solar neighborhood.  These independent relations unambiguously
determine the upper mass limit of $m_u=50 \pm 10$ \ms and the IMF
slope index of 1.3 -- 1.6 above 1 \msp.  This upper mass limit
corresponds to the mass beyond which stars end as black holes without
ejecting processed matter into the interstellar medium.
We also find that the IMF slope index below 0.5 \ms cannot be much 
shallower than 0.8.

\end{abstract}

%        --- stars: black hole}

\section{INTRODUCTION}

The initial mass function of stars (IMF) is a key ingredient in
modeling the evolution of galaxies.  The IMF shape together with the
lower and upper bounds of stellar mass largely influences the
resulting abundance pattern of heavy elements ejected through
supernovae explosions, the total amount of mass contained in stellar
remnants, and the integrated photometric properties of galaxies
(cf. Tinsley 1980).

As a conventional practice the IMF is assumed to be a time-invariant
mass spectrum having a power-law of the form $n(m)dm \propto
m^{-(1+x)}dm$ ($m_{\ell}\leq m \leq m_u$), where $n(m)dm$ is the
number of stars in the mass interval $m$ to $m+dm$.  The IMF is
usually derived from the observed luminosity function of stars, but
even the local IMF in the solar neighborhood is still controversial
ranging between Salpeter's (1955) IMF ($x$=1.35) and Scalo's (1986)
IMF ($x$=1.9) for $m$ \gtsim 1 \msp. The integrated photoionization
rates and colors of nearby disk galaxies are fitted by a Salpeter-like
shallower IMF (Kennicutt 1983; Kennicutt, Tamblyn, \& Congdon 1994;
Sommer-Larsen 1996).  Such a global IMF for disk galaxies could be
compared to the solar neighborhood IMF, because theoretical arguments
indicate that the IMF is originated from fragmentation of gas cloud in
a way nearly independent of local physics in the gas although the IMF
slope might change in starbursts (Silk 1977, 1995; Yoshii \& Saio
1985; Price \& Podsiadlowski 1995).

This controversial situation for determining the IMF slope holds also
for the Large Magellanic Cloud (LMC).  Through transformation of the
photometric stellar data to the theoretical H-R diagram, Hill, Madore,
\& Freedman (1994) have obtained a steep IMF ($x$ = 2), whereas
significantly shallower slope is derived by Massey et al. (1995) using
a similar method ($x$=1.3), by Hunter et al. (1995) from recent HST
observation ($x$ = 1.22), and by Will et al. (1995a,b) from the
stellar luminosity function ($x$ = 1.1 -- 1.2).  Such a difference in
the reported IMFs for either the solar neighborhood or the LMC
suggests that the previous methods can not enable a precise
determination of the IMF slope.

In addition the upper mass limit $m_u$ remains still undetermined.
The ratio $Y/Z$ ($Y$: helium, and $Z$: {\sl metals} more massive than
He) is a strongly decreasing function of progenitor mass. Then, the
relative change in helium to metal abundance $\Delta Y/\Delta Z$ from
the big bang to the present time is strongly dependent on $m_u$.
Maeder (1992, 1993), based on his stellar models with the empirical
mass loss function, found that the observed high ratio $\Delta
Y/\Delta Z$ in extragalactic H$\,$II regions (Walker \etal 1991;
Pagel et al. 1992) can only be matched with $m_u$ being as low as 20 --
25\msp.  This is significantly smaller than the estimate of $m_u\sim
40 - 80$\ms by van den Heuvel (1992) from the presence of black hole
X-ray binaries and X-ray pulsars in high mass X-ray binaries.

Recent accumulation of observed heavy-element abundances for the
solar-neighborhood stars offers an opportunity of using a new method
for the IMF determination based on the supernova nucleosynthesis
argument. In a supernova explosion, specific nucleosynthetic features
will be revealed, and the features averaged over an IMF are
incorporated in the gas from which stars of next generations are born.
The stars below 1\ms do not eject the synthesized heavy elements into
the interstellar medium, but confine the heavy elements produced by
massive stars when stars are formed. From a viewpoint of chemical
evolution, it does not matter how stars below 1 \ms are distributed.
The important point is therefore the total stellar mass of such
low-mass stars. In other words, we can assume a single slope IMF over
the whole mass range and can know the mass fraction below 1 \ms from
the combination of the lower mass limit $m_{\ell}$ and the IMF slope
$x$ which can reproduce the observed chemical properties.  

The mass of ejecta of elements like O, Ne, and Mg from Type II
supernovae (SNe II) increases strongly with progenitor mass.  On the
other hand, the mass of products from explosive burning Si, S, Ar, Ca,
and Fe remains close to constant as a function of progenitor mass or even
slightly decreases in the case of Fe (Nomoto \& Hashimoto 1988;
Thielemann, Nomoto, \& Hashimoto 1993; Hashimoto \etal 1993a, 1996).
Thus, the average ratios of O/Fe, Ne/Fe, and Mg/Fe, as they result
from a whole population of SNe II integrated over the IMF, will also
depend strongly on $m_u$ while a weaker dependence is expected for
Si/Fe through Ca/Fe.

\begin{deluxetable}{ccrccccc}
\footnotesize
\tablecaption{The chemical evolution models for the solar neighborhood}
\tablehead{
&\multicolumn{3}{c}{Input quantities at $T_G=15$Gyr} & & 
\multicolumn{3}{c}{Calculated metallicity yields} \\ 
\cline{2-8}
\colhead{$k$} & \colhead{$f_g(T_G)$} & \colhead{[O/H]$_g$} 
& \colhead{[Fe/H]$_g$} & & \colhead{$p_{\rm II,O}$} & 
\colhead{$p_{\rm II,Fe}$} & \colhead{$p_{\rm Ia,Fe}$} \\
}
\startdata
1 & 0.19  & --0.03 & 0.05 & & 8.03$\times10^{-3}$ & 3.12$\times10^{-4}$ 
& 7.76$\times10^{-4}$\\ 
  & 0.25 &   0.00 & 0.08 & & 9.54$\times10^{-3}$ & 3.71$\times10^{-4}$
& 9.56$\times10^{-4}$\\ 
2 & 0.19  & --0.03 & 0.05 & & 8.21$\times10^{-3}$ & 3.20$\times10^{-4}$
& 7.63$\times10^{-4}$\\ 
  & 0.25 &   0.00 & 0.08 & & 9.54$\times10^{-3}$ & 3.71$\times10^{-4}$
& 9.23$\times10^{-4}$\\ 
3 & 0.19  & --0.03 & 0.05 & & 8.42$\times10^{-3}$ & 3.28$\times10^{-4}$
& 7.66$\times10^{-4}$\\ 
  & 0.25 &   0.00 & 0.08 & & 9.66$\times10^{-3}$ & 3.76$\times10^{-4}$
& 9.19$\times10^{-4}$\\ 

\enddata

\tablenotetext{}{Note. --- 
$f_g(T_G)$ is the present mass fraction of the gas. [O/H]$_g$ and [Fe/H]$_g$ 
are the present abundances of oxygen and iron in the gas, respectively.
}

\tablenotetext{}{Note. --- The parameters used in common are $t_{\rm
Ia}$ = 1.5 Gyr (the lifetime of Type Ia supernova progenitors),
$t_{\rm in}$ = 5 Gyr (the timescale of the gas infall), and
[O/Fe]$_{\rm II}$ = +0.4 (the oxygen to iron abundance ratio for metal-poor
population II stars).  
}

\end{deluxetable}

The O/Fe abundance ratio of nearby stars is among the best observed
ones of these ratios (e.g., Barbuy 1988; Barbuy \& Erderlyi-Mendes
1989; Gratton 1991; Nissen \& Edvardsson 1992; Edvardsson \etal 1993).
Yoshii, Tsujimoto, \& Nomoto (1996, hereafter YTN) used the standard
chemical evolution model and determined the O and Fe yields from SNe
II. These empirical yields are compared with the updated supernova
nucleosyntheis yields (Hashimoto, Nomoto, \& Shigeyama 1989, 1996;
Thielemann, Nomoto, \& Hashimoto 1996) to derive the relation between
the lower and upper mass limits ($m_{\ell}$, $m_u$) and the
IMF slope $x$. [Wang \& Silk (1993) derived the $m_{\ell}-x$
relation using a fixed value of $m_u$ = 60\ms and the averaged O yield
over those given by Arnett (1978), Woosley \& Weaver (1986), Maeder
(1992), and Thielemann, Nomoto, \& Hashimoto (1994).]  In order to
determine a unique combination of ($m_{\ell}$, $m_u$, $x$),
we need another relation among these IMF parameters.

We note that the stellar mass to light $M/L$ ratio provides an
independent constraint on the IMF.  As in the case of chemical
evolution, we can assume a single slope IMF in calculating this ratio,
because the luminosity from stars below 1\ms is negligible and
therefore how stars below 1\ms are distributed is not important.
Moreover, as far as an IMF decreases towards high masses, the $M/L$
ratio depends sensitively on $x$ and $m_{\ell}$, but only very weakly
on $m_u$ if $m_u\gtsim 30$\msp.  Thus, we obtain the $m_{\ell}-x$
relation which reproduces the present value of $M/L$ in the solar
neighborhood.

In this paper we construct the chemical evolution model which
reproduces the major observational features in the solar neighborhood
(\S 2).  Thereby we derive the $m_{\ell}-x$ and $m_u-x$
relations from the oxygen to iron abundance ratios of metal-poor stars
(\S 3) and the $m_{\ell}-x$ relation from the stellar mass
to light ratio (\S 4).  These relations are used to determine
$m_{\ell}$, $m_u$, and $x$ for the local IMF in the solar
neighborhood (\S 5).  The result of the paper is discussed in \S 6.

\section{THE CHEMICAL EVOLUTION MODEL}

\begin{figure*}
\includegraphics{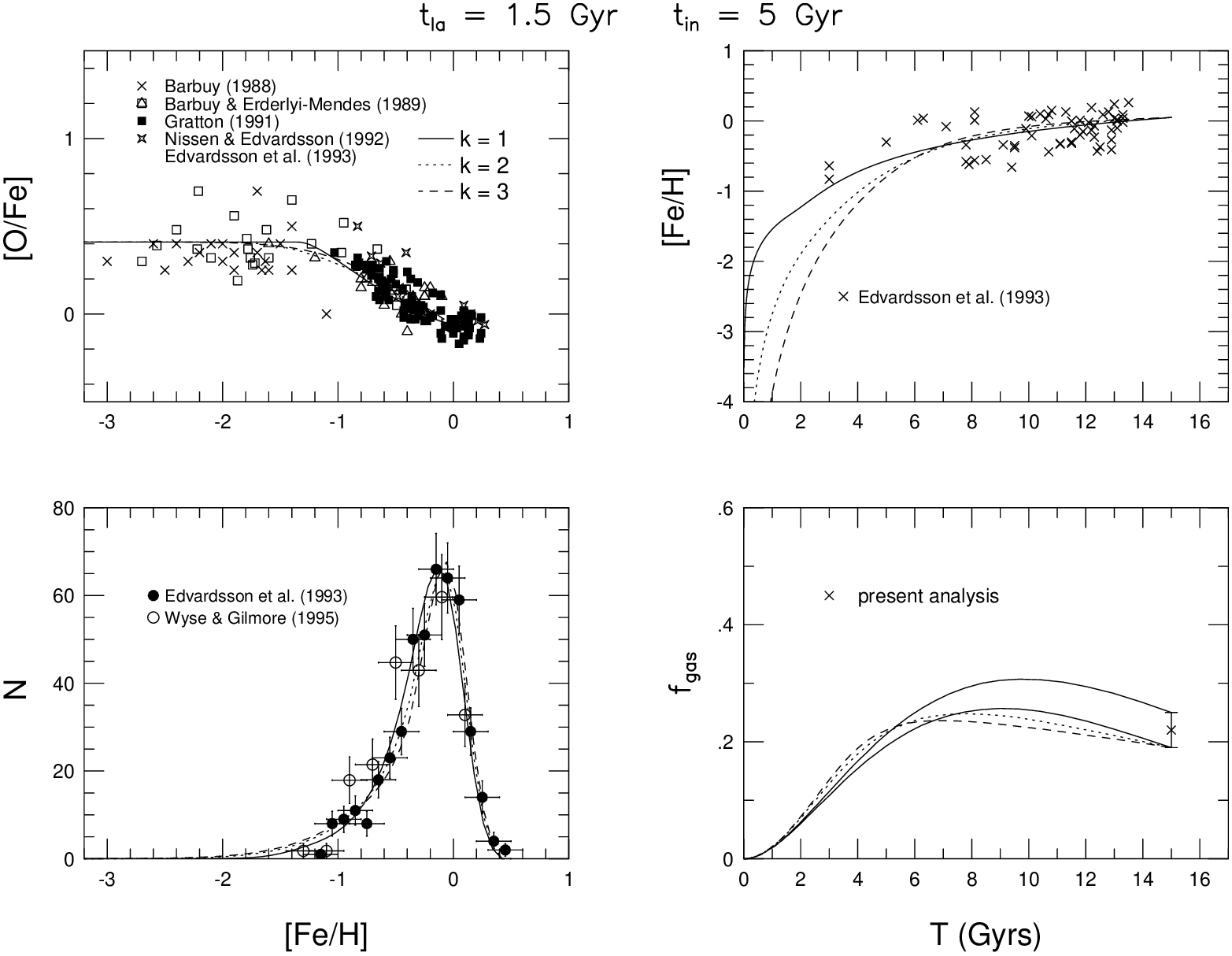}
\vspace*{15cm}
%\begin{figure*}
%\epsfxsize=470pt
%\epsfysize=700pt
%\angle=270
%\epsfbox{../psdata/fig1.eps}
%\psfig{psfile=../psdata/fig1.ps voffset=-450 vscale=0.7 hscale=0.7 
%hoffset=-20}
%\vspace*{15cm}
\caption{ 
Features of chemical quantities in the solar neighborhood.
Shown are the models with $t_{\rm Ia}=1.5$Gyr, $t_{\rm in}=5$Gyr and
$f_g(T_G)$=0.19 for $k=1-3$.  The symbols represent the data taken from
various papers.  The model metallicity distributions and observed one
by Wyse \& Gilmore (1995) are normalized to coincide with the total
number of the sample stars used by Edvardsson et al. (1993).  The
models for $f_g(T_G)$=0.25 are indistinguishable from those for
$f_g(T_G)=0.19$ and are therefore not shown except in the lower right
panel.  }
\label{fig-1}
\end{figure*}

SNe II produce most of oxygen after a short lifetime $t_{\rm II}\sim
10^{6-7}$ yrs due to their massive progenitor stars, while SNe Ia
produce most of iron delayed by a considerably longer lifetime $t_{\rm
Ia}\sim 1$ Gyrs of their less massive progenitor stars.  We employ an
instantaneous recycling approximation $t_{\rm II}\ll t$ only for SN II
progenitors, and two nucleosynthesis approximations such as (i)
$y_{\rm II,O}/y_{\rm II,Fe}=$ $(Z_{\rm O}/Z_{\rm Fe})_\odot 10^{\rm
[O/Fe]_{\rm II}}$ = constant where [O/Fe]$_{\rm II}$=+0.4 for
[Fe/H]$<$--1 (Gratton 1991), and (ii) $y_{\rm Ia,O}/y_{\rm II,O}\ll 1$
(Tsujimoto et al. 1995).  Hereafter we use the effective yield defined
as $p_i\equiv y_i/\alpha$ where $y_i$ is the mass of ejected heavy
element $i$ and $\alpha$ is the mass fraction locked up in stellar
remnants and low mass stars in each generation of stars. [We note that
this fraction $\alpha$ and a rate coefficient $\nu$ of star formation
always appear as their product in the chemical evolution equations.]
The use of these approximations in the chemical evolution enables an
empirical estimate of $p_{\rm II,O}$, $p_{\rm II,Fe}$, and $p_{\rm
Ia,Fe}$ without knowing the IMF in advance.

Following the procedure described in YTN, we construct a simplified
model of chemical evolution allowing for material infall from outside
the solar neighborhood.  The chemical evolution equations describing
time variation of gas fraction $f_g$ and the abundances of oxygen
$Z_{\rm O}$ and iron $Z_{\rm Fe}$ are then solved under the boundary
condition that these quantities at $T_G$=15 Gyrs should coincide with
those observed in the present disk.  We assume that the star formation
rate is proportional to some power $k$ of gas fraction $C(t)=\nu
[f_g(t)]^k$ and that the infall rate has a modified exponential form
with a timescale $t_{\rm in}$.  For a possible range of $k=1-3$ we fix
$t_{\rm Ia}=1.5$ Gyr and $t_{\rm in}=5$ Gyr, because both the
[O/Fe]--[Fe/H] diagram and the [Fe/H] abundance distribution function
of long-lived stars are well reproduced by this combination of $t_{\rm
Ia}$ and $t_{\rm in}$ (YTN).
   
For the present gas fraction, we set $f_g(T_G)=0.19$ or 0.25 assuming
that the surface mass density of the gas component near the sun is 10
or 11.5 \ms${\rm pc}^{-2}$, compared with the total mass
density of 54 or 46 \ms${\rm pc}^{-2}$, respectively (see \S 4).  For
the present abundances of oxygen and iron, we set ([O/H]$_g$,
[Fe/H]$_g$)=(--0.03, +0.05) for $f_g(T_G)$=0.19, and (0.00, +0.08) for
$f_g(T_G)$=0.25, which are adjusted to give the best agreement with
the observed chemical properties near the sun.

Using the above inputs in the chemical evolution equations, we have
derived the values of $p_{\rm II,O}$, $p_{\rm II,Fe}$, and $p_{\rm
Ia,Fe}$.  The results are tabulated in Table 1 and are shown in Figure
1.  The models for $k=1-3$ in the case of $f_g(T_G)$=0.19 are shown in
the [O/Fe]--[Fe/H] diagram (the upper left panel), the [Fe/H]
distribution function of long-lived disk stars (the lower left panel),
the age--metallicity relation (the upper right panel), and the
evolutionary behavior of gas fraction (the lower right panel).  The
data taken from various papers are shown by symbols.  The models shown
in the lower left panel are normalized to coincide with the total
number of the sample stars used by Edvardsson et al. (1993).  We note
that the models for $f_g(T_G)$=0.25 are indistinguishable from those
for $f_g(T_G)$=0.19 in the first three panels and are therefore not
shown in these panels.  In the last panel, however, the model for
$k=1$ and $f_g(T_G)$=0.25 is also shown for reference.

The empirical oxygen yield $p_{\rm II,O}$ and the star formation rate
$C(t)$ derived here will be used to determine the IMF in the following
sections.
 
The empirical oxygen yield $p_{\rm II,O}$ and the star formation rate
$C(t)$ derived here will be used to determine the IMF in the following
sections.
 
\section{APPROACH FROM NUCLEOSYNTHESIS ARGUMENT}

\begin{figure*}
\includegraphics{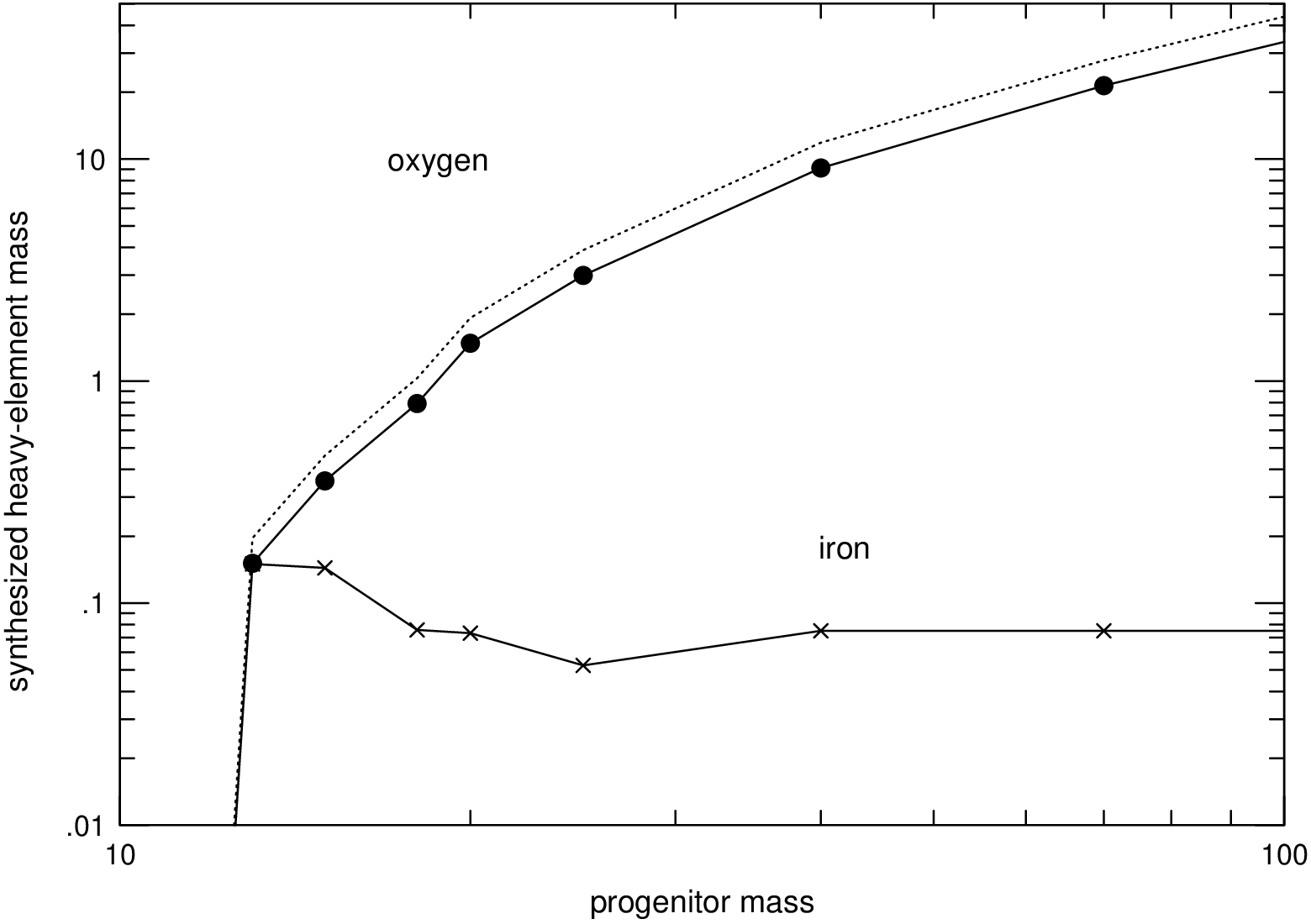}
%\epsfxsize=200pt
%\epsfysize=100pt
%\epsfbox{../psdata/fig2n.eps}
\vspace{8.2cm}
\caption{
Synthesized and ejected oxygen and iron masses from Type
II supernovae as a function of the mass of their progenitor
main-sequence stars which are assumed to have solar abundances.  The
masses are in units of \msp.  Considering an uncertainty of the oxygen
products we show two cases such as the standard oxygen product
obtained by our nucleosynthesis calcultaions (solid line) and the
high-O case in which the standard oxygen products for each
main-sequence mass are multiplied by a factor of 1.3 (dotted line).
}
\label{fig-2}
\end{figure*}

Comparing the empirical O and Fe yields with the updated theoretical
supernova yields, we can derive the lower and upper mass limits
($m_{\ell}, m_u$) as a function of the IMF slope index $x$.  These IMF
parameters are subject to the following two relations:

$$
\frac{p_{\rm II,O}}{p_{\rm II,Fe}}=
\frac{\displaystyle \int^{m_u}_{10M_\odot} M_{\rm II,O}(m) n(m) dm}
{\displaystyle \int^{m_u}_{10M_\odot} M_{\rm II,Fe}(m) n(m) dm}
\ \ , \eqno{(1)}
$$

and

$$ 
p_{\rm II,O} =
\frac{\displaystyle \int^{m_u}_{10M_\odot} M_{\rm II,O}(m) n(m) dm}
{\displaystyle \int^{m_u}_{m_{\ell}} M_{\rm rem}(m) n(m) dm}
\ \ . \eqno{(2)}
$$

The ejected oxygen and iron masses ($M_{\rm II,O}$, $M_{\rm II,Fe}$)
from SNe II are taken from explosive nucleosynthesis calculations for
massive stars with $m=10 - 70M_{\odot}$ by Hashimoto et al. (1989) and
Thielemann \etal (1990, 1996).  The nucleosynthesis contribution from
$m=8 - 10M_{\odot}$ stars is assumed to be negligible (Nomoto 1984,
1987; Hashimoto, Iwamoto, \& Nomoto 1993b).  These ejected masses for
each main-sequence stars are tabulated in Tsujimoto et al. (1995) and
are shown in Figure 2.  It is seen from this figure that the produced
oxygen increases steeply as the stellar mass increases.

The calculated oxygen mass is subject to the combined uncertainties
involved in the $^{12}$C($\alpha$, $\gamma$)$^{16}$O rate and
convective overshooting near the end of core He burning.  The higher
rate and smaller overshooting result in the smaller C/O ratio after He
burning and thus smaller Ne/O and Mg/O ratios after C burning.  The
range of combined uncertainties is constrained by the requirement that
the resultant SN II yields reproduce the solar ratios of Ne/O and
Mg/O.  The allowed range of oxygen masses thus obtained lies between
our standard masses and those by Woosley \& Weaver (1995), where the
latter calculation produces $\sim$ 1.3 times larger oxygen masses than
our standard case.
Taking into account the above uncertainty of the oxygen product, we
calculate two cases for the oxygen product, that is, the standard
oxygen product obtained by our nucleosynthesis calculations (solid
line) and the high-O case in which we multiply the standard oxygen
products for each main-sequence mass by a factor of 1.3 (dotted line).

There also exists an uncertainty concerning the iron product, but this
uncertainty is relatively small as far as an IMF decreases towards
high masses, because (1) the dependence of the ejected iron mass on
the progenitor mass is very weak and therefore (2) the integrated
iron product is mainly determined by a lower mass range 10 -- 20 \ms
as constrained by recent observations of Type II, Ib/c supernovae (SN
1983N, SN 1987A, SN 1990E, SN 1993J, SN 1994I; Nomoto, Iwamoto, \&
Suzuki 1995).  Therefore we take into account only the uncertainty of
the oxygen product.

The remnant masses $M_{\rm rem}$ are taken as the masses of white
dwarfs for the initial low-mass stars, and as the masses of neutron
stars for the initial massive stars.  The initial-final mass relations
for white dwarfs and neutron stars are described in YTN. ( For the 70
\ms star, $M_{\rm rem}$ = 1.57 \msp, which is not given in YTN.) As
for the lower mass limit below which stars do not experience the mass
loss, we set $m_{{\rm wd},l}$=0.7 $M_\odot$ and therefore $M_{\rm
rem}$=$m$ below $m$=$m_{{\rm wd},l}$.

Using the empirical O/Fe yield ratio in the LHS and the
nucleosynthesis product masses in the RHS, we derive the upper mass
limit $m_u$ as a function of the IMF slope index $x$ from eq.[1].  The
result is shown in the upper panel of Figure 3, where solid and dotted
lines represent the standard and high-O cases of the oxygen product,
respectively.  For the standard case a Salpeter IMF ($x=1.35$)
corresponds to $m_u=50$ \ms and a Scalo IMF ($x=1.9$) to $m_u=80$
\msp, whereas these are $m_u=40$ and 50\ms for the high-O case.  We
note that an uncertainty of $m_u$ arising from the uncertainty in the
oxygen products becomes larger for a steeper IMF.

Similarly using the obtained $m_u - x$ relation in eq.[2], we derive
the lower mass limit $m_{\ell}$ as a function of $x$.  Since we assume
a single IMF slope over the whole mass range, the lower mass limit
derived here might not represent the real $m_{\ell}$ if the IMF
slope changes below 1 \msp. The derived $m_{\ell}$ is a parameter
to determine the total stellar mass for $m$ $<$ 1 \msp, as mentioned
in \S 1.  The result of the $m_{\ell} - x$ relation is shown in the
lower panel of Figure 3 where two thin lines correspond to those for
the lowest and highest values of $p_{\rm II,O}$ in Table 1.  Contrary
to the $m_u - x$ relation, different choices of $M_{\rm II,O}$ hardly
change the $m_{\ell} - x$ relation because the IMF steeply decreases
towards high masses so that the RHS of eq.[2] is not very sensitive to
$m_u$.  From the obtained $m_u - x$ and $m_{\ell} - x$ relations, the
mass range covers 0.04 \ms -- 50 \ms for a Salpeter IMF and 0.4 \ms --
80 \ms for a Scalo IMF.

In the present analysis we do not include a possible metallicity
dependence of the nucleosynthesis yields.  Oxygen and Fe are primary
yields and do not depend on seed abundances from previous stellar
populations.  They could only be affected by a metallicity dependence
of stellar structure, which are not expected to be large (Woosley \&
Weaver 1982).  More significant is the influence of metallicity
dependent stellar wind losses.  If the O output from SNe II is caused
almost entirely by yields from hydrostatic stellar evolution, mass
loss would cause a drastic change of O/Fe with respect to calculations
which neglect this mass loss (Woosley, Langer, \& Weaver 1993).  In
particular, the oxygen yield integrated over an appropriate IMF
becomes smaller than that without mass loss, but the difference is not
very significant (Wang \& Silk 1993; Prantzos, Vangioni-Flam, \&
Chauveau 1994).  This effect of mass loss sets only in for progenitor
masses of 40--60 $M_\odot$ with metallicities of [Fe/H] $\geq$ -- 0.4
(Langer 1989; Maeder 1990,1992; Charbonnel et al. 1993; Schaerer et
al. 1993).  Therefore the $m_u - x$ relation stays unaltered because
our analysis of the O/Fe ratio is based on the metal-poor stars before
such metallicities are attained.  On the other hand, given the
empirical value of $P_{\rm II,O}$, the induced reduction in the oxygen
yield in the numerator of eq.[2] requires lowering the stellar remnant
mass in the denominator, which corresponds to an upward shift of the
$m_{\ell} - x$ relation in Figure 3.

\section{APPROACH FROM THE $M/L$ ARGUMENT}

The total stellar mass $M_{\ast}$ is very sensitive to the lower mass
limit $m_{\ell}$, and the total stellar luminosity $L_{\ast}$ is
sensitive to the IMF slope $x$ rather than $m_{\ell}$.  For the
decreasing IMFs with increasing stellar mass, the upper mass limit
$m_u$ hardly affects either $M_{\ast}$ or $L_{\ast}$ as far as $m_u$
\gtsim 30 \ms because such massive stars are very few in number and
have very short lifetimes.  This indicates that the $(M/L)_{\ast}$
ratio is essentially determined by a combination of $m_{\ell}$ and
$x$.

Using the rate of star formation $C(t)$ which reproduces the 
solar-neighborhood chemical quantities (\S 2), we calculate the
present stellar mass $M_{\ast}(T_G)$ and luminosity $L_{V\ast}(T_G)$
according to the equations below:
$$
\hspace{-0.cm} M_{\ast}(T_G) = \int^{T_G}_{t_{m_u}} \hspace{-0.1cm}
dt C(T_G - t)
$$
\vspace{-0.2cm}
$$
\times
\left\{ \int^
{m_t}_{{\rm min}(m_{\ell},m_t)} \hspace{-0.9cm} m n(m) dm   
+ \int^{m_u}_{m_t} \hspace{-0.2cm}
M_{\rm rem}(m) n(m) dm \right\} \ \ , \eqno{(3)}
$$
and
$$
\hspace{-0.cm} L_{V\ast}(T_G) = \int^{T_G}_{t_{m_u}} \hspace{-0.1cm}
dt C(T_G - t) \int^{m_u}_{{\rm min}(m_{\ell},m_t)}
\hspace{-0.9cm} l_V(m) n(m) dm \ \ . \eqno{(4)}
$$
The turnoff mass $m_t$ corresponding to an age of $t$ is given
by (Renzini \& Buzzoni 1986)
$$
\hspace{-1.cm}
\log[m_t/M_\odot] = 0.0558[\log (t/{\rm yrs})]^2 
$$
\vspace{-0.2cm}
$$
\hspace{2 cm} - 1.338\log (t/{\rm yrs}) 
+ 7.764 \ \ , \eqno{(5)}
$$
and $t_{m_u}$ is the lifetime of stars with mass $m_u$.  In eq.[3] the
first and second terms in the braces denote the contribution from
main-sequence stars and stellar remnants, respectively.  The stellar
mass-luminosity relation $l_V(m)$ in the $V$ band, which is used in
eq.[4], is taken from the theoretical solar-abundance $m$ -- $M_V$
relation compiled by Tinsley (1980).  This relation is consistent with
the recent theoretical relation by Tout et al. (1996) and agrees with
the empirical relation by Scalo (1986). In eq.[4] we do not consider a
possible metallicity effect in the $m$ -- $M_V$ relation nor the
luminosity from giant stars.  This is because the majority of
long-lived disk stars of our concern have metallicities as high as
[Fe/H] $\sim$ -- 0.1 (see the lower left panel of Figure 1) and there
is a large uncertainty in estimating the luminosity of giant stars.

\begin{figure*}
\includegraphics{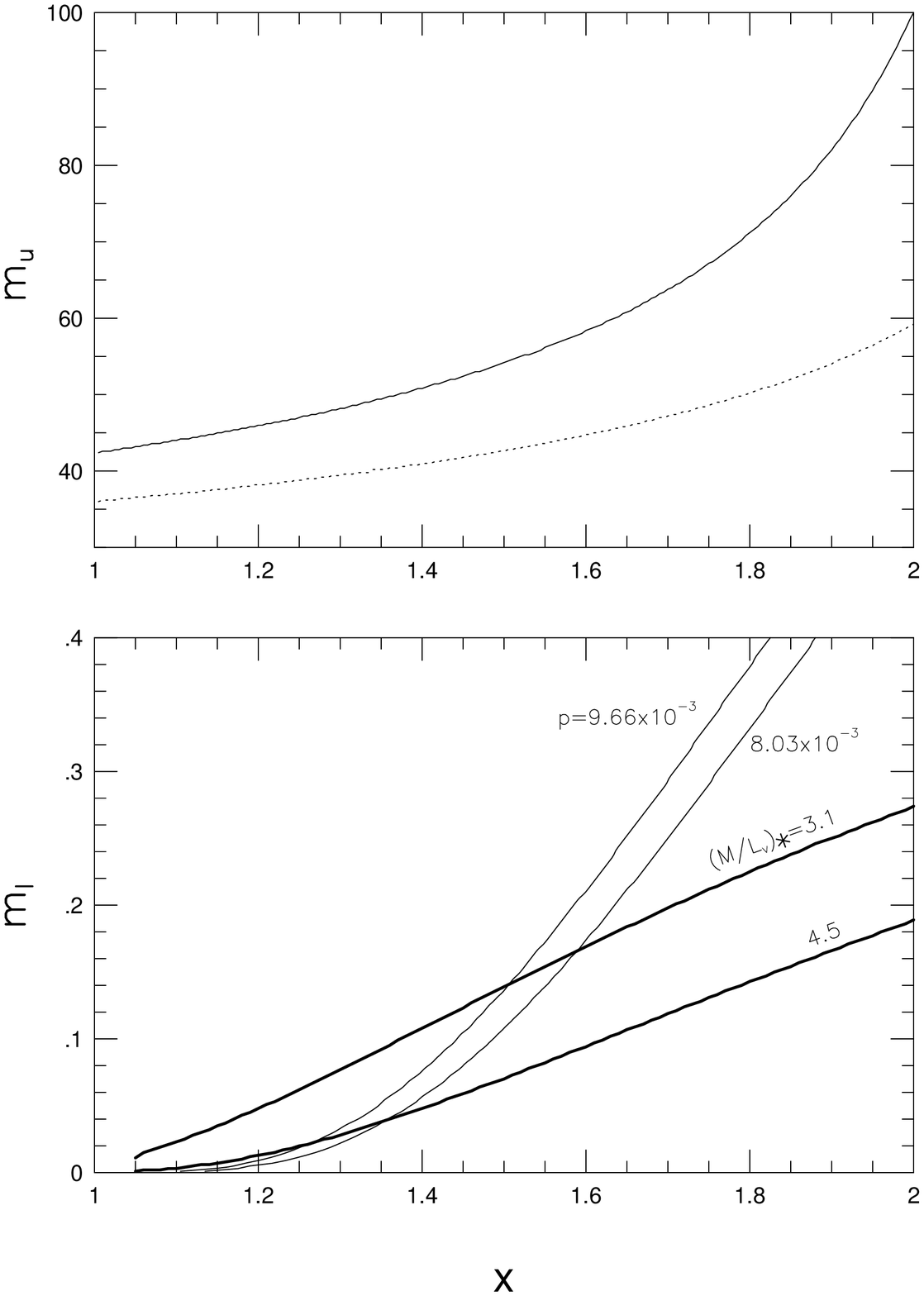}
\vspace*{17cm}
\caption{
Upper and lower mass limits $(m_u, m_{\ell})$ in units of
\ms as a function of the IMF slope index $x$.  In the upper panel
shown are the $m_u - x$ relations for the standard (thick line) and
high-O (dotted line) cases of the oxygen product.  In the lower panel
shown are the $m_{\ell} - x$ relations based on the nucleosynthesis
argument (thin lines) and the mass-to-light ratio argument (thick
lines). 
}
\label{fig-3}
\end{figure*}

\begin{figure*}
\includegraphics{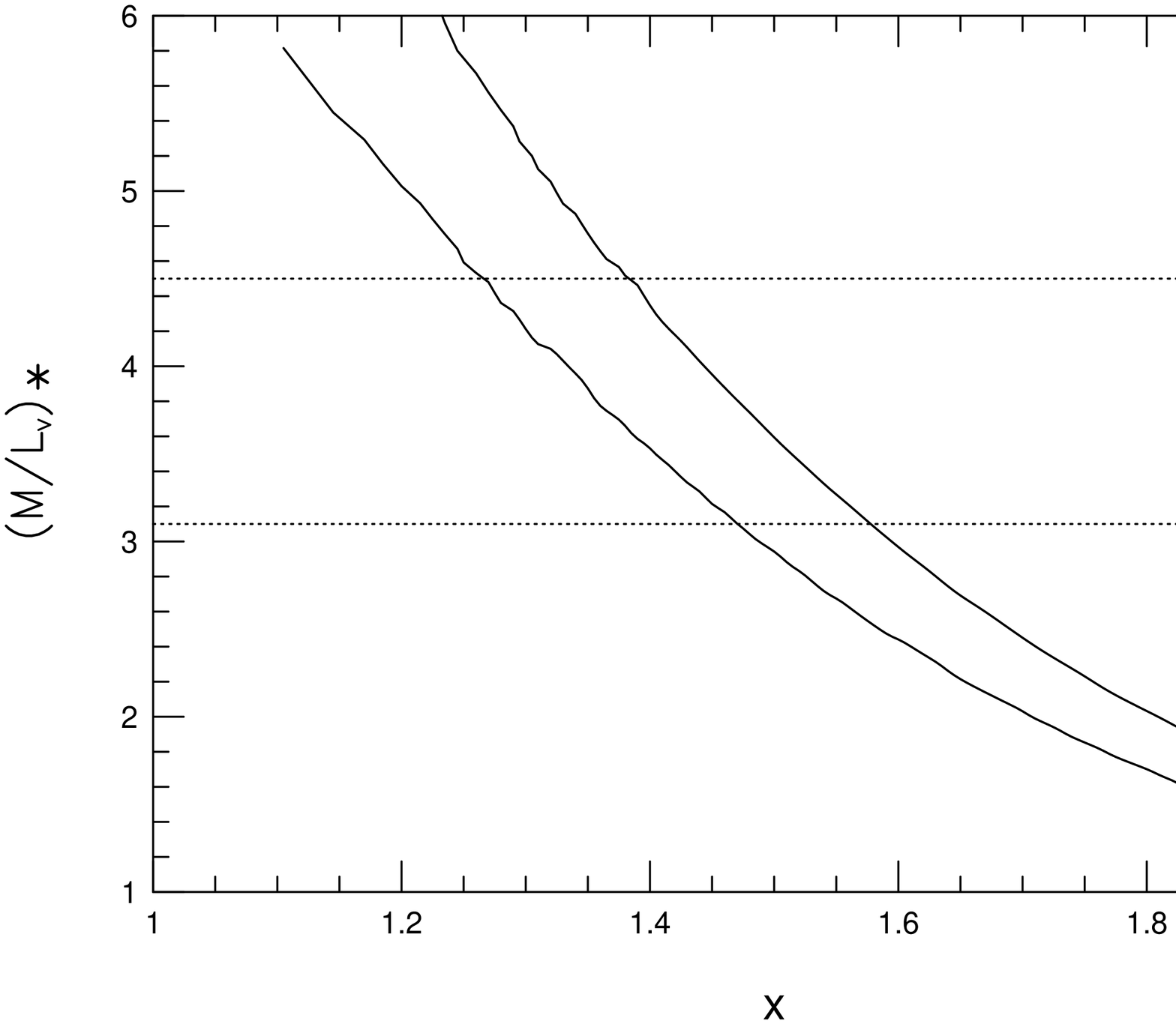}
\vspace{8.2cm}
\caption{
Mass to visual light ratio $(M/L_V)_\ast$ for disk stars
in unit of $(M/L_V)_\odot$ as a function of the IMF slope index $x$
for different oxygen yields (upper thick line for
$p=8.03\times10^{-3}$ and lower thick line for $p=9.66\times10^{-3}$).
The observational bounds of $(M/L_V)_\ast$=3.1--4.5 are shown by the
dotted lines. 
}
\label{fig-4}
\end{figure*}

The observed $(M/L_V)_{\ast}$ ratio, which will be compared with the
calculated ratio, is obtained from the surface mass density of nearby
disk stars and their surface luminosity density of only main-sequence
stars. Since the dynamical mass in the solar neighborhood is a sum of
masses of stars and gas (Kuijken \& Gilmore 1989,1991), the stellar
contribution is evaluated from more direct measurements of the other
two, e.g., subtraction of the gas mass from the dynamical mass.
Various estimates of the dynamical mass from analyses of vertical
motion of disk stars are converged to give $\Sigma_{\rm dyn}$=46 -- 54
\ms${\rm pc}^{-2}$ (Kuijken \& Gilmore 1989,1991; Gould 1990; Flynn \&
Fuchs 1994).  Much larger values of 67 and 84 \ms${\rm pc}^{-2}$
(Bahcall 1984; Bahcall, Flynn, \& Gould 1992) are attributed to their
inadequate choice of tracer stars and restrictive analysis technique
(Kuijken 1995).  The mass of gas is investigated in detail by Kuijken
\& Gilmore (1989), and their estimate leads to $\Sigma_{\rm gas}$=13
$\pm$ 3 \ms${\rm pc}^{-2}$.  Straightforward application of this
estimate of $\Sigma_{\rm gas}$ gives the gas fraction of 19 -- 35 \%,
the upper bound of which is unrealistically high. Indeed Young \&
Scoville (1991) study the gas fraction for all type of spiral galaxies
and report that Sb galaxies like our Galaxy have a definitive upper
bound of 25 \%. Accordingly we adopt the gas fraction of 19 -- 25 \%
which corresponds to $\Sigma_\ast$=35 -- 44 \ms${\rm pc}^{-2}$ for the
surface mass density of nearby disk stars. We note that stellar
remnants, brown dwarfs and giant stars are not subtracted out of the
above estimate of $\Sigma_\ast$.

Star count analyses in the direction perpendicular to the Galactic
disk provide an estimate of the surface luminosity density of all
stars in the $V$ band such as $\Sigma_V$=23.3 \ls${\rm pc}^{-2}$
(Bahcall 1984), 23.8 \ls${\rm pc}^{-2}$ (van der Kruit 1986), and 25.3
\ls${\rm pc}^{-2}$ (Yoshii, Ishida, \& Stobie 1987).  These values are
consistent with the value of 24 \ls${\rm pc}^{-2}$, which is evaluated
from $2.4\mu m$ observations assuming $V-K$ = 3.2 (Fazio, Dame, \&
Kent 1990).  By subtracting a contribution from giant stars, these
authors have also estimated the surface luminosity density of only
main-sequence stars as 9.7 \ls${\rm pc}^{-2}$ (Bahcall 1984), 10.2
\ls${\rm pc}^{-2}$ (van der Kruit 1986), and 11.4 \ls${\rm pc}^{-2}$
(Yoshii, Ishida, \& Stobie 1987), respectively, leading to 9.7 -- 11.4
\ls${\rm pc}^{-2}$ for a reliable range.

We therefore conclude that the observed $(M/L_V)_{\ast}$ ratio in the
solar neighborhood is constrained to be 3.1 -- 4.5 $(M/L_V)_\odot$.
Using this empirical ratio and theoretical inputs in eqs.[3] and [4],
we obtain the $m_{\ell}-x$ relation as shown by thick lines in the
lower panel of Figure 3.  In eq.[3] we tentatively use $m_u$ = 50
\msp, but different choices hardly change the result.  The allowable
range of $m_{\ell}-x$ relation is confined between two lines
corresponding to $(M/L_V)_{\ast}$ = 3.1 and 4.5 $(M/L_V)_\odot$.

\section{THE IMF DETERMINATION}

We have obtained two $m_{\ell} - x$ relations which are
independent of each other; one relation shown by thin lines in the
lower panel of Figure 3 is from the nucleosynthesis argument (\S 3),
and another relation by thick lines is from the mass-to-light ratio
argument (\S 4).  It is evident that two allowable regions overlaps in
a narrow range of $x$ = 1.3 -- 1.6.

\begin{table}[h]
\bc
\caption{Derived parameters for the solar
neighborhood IMF} 
\begin{tabular}{cccccccc}
\hline \hline
\multicolumn{8}{c}{single slope} \\
\hline 
$x$ & & $m_u$ & & $m_{\ell}$ & & $f_{\rm BD}$ & \\ 
& & (STD/high-O) & & & & \\
\hline
1.3 & & 48/40 & & 0.03 & & 0.29 & \\ 
1.4 & & 51/41 & & 0.07 & & 0.06& \\ 
1.5 & & 54/43 & & 0.12 & & 0.00 & \\ 
1.6 & & 59/45 & & 0.16 & & 0.00 & \\ 
\hline \hline 
\multicolumn{8}{c}{double slope} \\ \hline
\multicolumn{2}{c}{$m \geq 1 M_{\odot}$} & 
\multicolumn{6}{c}{$m < 1 M_{\odot}$} \\ 
\cline{1-2} 
\cline{3-8} 
$x$ & $m_u$ & $x$ & & $m_{\ell}$ 
& & $f_{\rm BD}$ & \\
\hline 
1.3 & 48 & 1.2 & & 0.02 & & 0.34 & \\ 
& & 1.1 & & 0.01 & & 0.39 & \\ 
1.4 & 51 & 1.2 & & 0.04 & & 0.22 & \\ 
& & 1.1 & & 0.02 & & 0.28 & \\ 
1.5 & 54 & 1.2 & & 0.05 & & 0.09 & \\ 
& & 1.1 & & 0.03 & & 0.17 & \\ 
1.6 & 59 & 1.2 & & 0.08 & & 0.01 & \\ 
& & 1.1 & & 0.06 & & 0.10 & \\ 
\hline 
\multicolumn{2}{c}{$m \geq 0.5 M_{\odot}$} &
\multicolumn{6}{c}{$m < 0.5 M_{\odot}$} \\ 
\cline{1-2}
\cline{3-8} 
$x$ & $m_u$ & $x$ & & $m_{\ell}$ & & $f_{\rm BD}$ & \\ 
\hline 
1.3 & 48 & 0.85 & & 0.0004 & & 0.41 & \\ 
& & 1.0 & & 0.006 & & 0.38 & \\ 
1.4 & 51 & 0.85 & & 0.01 & & 0.27 & \\ 
& & 1.0 & & 0.03 & & 0.23 & \\ 
1.5 & 54 & 0.85 & & 0.04 & & 0.13 & \\ 
& & 1.0 & & 0.06 & & 0.07 & \\ 
1.6 & 59 & 0.85 & & 0.08 & & 0.01 & \\ 
& & 1.0 & & 0.10 & & 0.00 & \\ 
\hline
\end{tabular}
\vskip 0.2pc
\ec
\hskip 1pc \small{Note. --- The upper and lower mass limits $m_u$ and
$m_{\ell}$ are in units of solar mass \msp. The symbol $f_{\rm BD}$ 
stands for the mass fraction of brown dwarfs below the hydrogen-burning 
minimum mass of 0.08 \ms but more massive than $m_{\ell}$.
}
\end{table}

Instead of displaying two independent $m_{\ell} - x$
relations separately, we substitute the nucleosynthesis $m_u-x$ and
$m_{\ell} - x$ relations in eqs.[3] and [4] to calculate
$(M/L_V)_{\ast}$ as a function of only $x$. The resulting
$(M/L_V)_{\ast}-x$ relations for different choices of empirical oxygen
yield are shown by thin lines in Figure 4.  Considering the
uncertainty of this relation and an observed range of $(M/L_V)_{\ast}$
= 3.1 and 4.5 $(M/L_V)_\odot$ (dotted lines), we can obtain $x=1.3 -
1.6$ which gives the best estimate of $m_u=50 \pm 10$ \ms from
Figure 3.

If we adopt the metallicity-dependent yield, the $m_{\ell} - x$
relations, shown by thin lines in the lower panel in Figure 3 (\S 3),  
move upwards, leading to a slightly shallower IMF.

\section{DISCUSSION}

The IMF slope derived here is consistent with the slope $x$ = 1.35 --
1.5 inferred from the photoionization properties of nearby galaxies
(Kennicutt 1983; Kennicutt et al. 1994; Sommer-Larsen 1996).  Much
steeper IMFs suggested by Scalo (1986) are ruled out because they fail
to reproduce the O/Fe ratio of metal-poor stars and the
$(M/L_V)_{\ast}$ ratio in the solar neighborhood simultaneously.

The upper mass limit of stars corresponds to the maximum supernova
progenitor mass beyond which the stars form black holes without
ejecting nucleosynthesis products into the interstellar medium rather
than neutron stars, the formation of which is associated with the
ejection of substantial amounts of metal enriched material.  While the
details of the supernova mechanism still await a final explanation, an
observational constraint from the O/Fe ratio of metal-poor stars allow
to put a limit of $m_u=50 \pm 10$ \ms on the progenitor mass
where the transition from neutron star to black hole remnants occurs.
This agrees with the estimate of $m_u=40 - 80$\ms by van den Heuvel \&
Habets (1984) and van den Heuvel (1992) from the presence of black
hole X-ray binaries and X-ray pulsars in high mass X-ray binaries.

The IMF, having a single slope over the whole mass range, extends down
to $m_{\ell}=0.03 - 0.16$\ms which is comparable to the
hydrogen-burning minimum mass ($m\approx 0.08$\msp).  It is very
difficult to determine the IMF below 1\ms because it is hard to
constrain the luminosity function and there are large uncertainties in
the mass--luminosity relation.  
%There might exist the turnover of the
%IMF towards less massive stars and the IMF slope below 1\ms might be
%shallower or bimodal 
The IMF slope might be shallower below 1 \ms and 0.5 \ms 
(Kroupa et al. 1993).  Taking this possibility into accout 
we first consider that the IMF slope $x$ below 1\ms could be shallower
than that for the massive part of the IMF, provided that the mass
fraction below 1\ms is kept unchanged.  Since $m_{\ell}$ for $x \ltsim
1$ becomes smaller than the theoretical estimate of the minimum Jeans
mass 0.007 -- 0.01 \ms (Low \& Lynden-Bell 1976; Silk 1977), we
restrict ourselves to $x=1.1-1.2$ and estimate how much mass is
contained in nonradiating objects (brown dwarfs).  For reference the
mass fraction $f_{\rm BD}$ of such objects below 0.08\ms but more
massive than $m_{\ell}$ is tabulated in Table 2. Secondly we consider
an IMF that is shallower below $m$ = 0.5 \msp, which
is suggested from the starcount analysis by Kroupa et al. (1993).  The
result for the IMF slope $x$ = 0.85 -- 1.0 below $m$ = 0.5 \ms is also
tabulated in Table 2.  It is suggested from this table that brown
dwarfs may exist in the solar neighborhood (see Carr 1994 for a recent
review), but we can not estimate their exact mass fraction unless the
IMF slope below 1 \ms is known.

An independent estimate of the mass fraction of brown dwarfs is
possible knowing that the dynamical mass is the sum of the gas mass,
luminous stars, and brown dwarfs.  We here use $\Sigma_{\rm dyn}$ =
$50 \pm 4$ \ms${\rm pc}^{-2}$ (Kuijken \& Gilmore 1989,1991; Gould
1990; Flynn \& Fuchs 1994), $\Sigma_{\rm gas}$ = $13 \pm 3$ \ms${\rm
pc}^{-2}$ (Kuijken \& Gilmore 1989), and $\Sigma_*^{\rm lum}$=$26 \pm
2$ \ms${\rm pc}^{-2}$ (Bahcall \& Soneira 1980; Yoshii et al. 1987;
M\'{e}ra, Chabrier, \& Baraffe 1996), where $\Sigma_*^{\rm lum}$
stands for the stellar mass excluding the contribution from brown
dwarfs. We obtain the mass fraction of brown dwarfs as 30 $\pm 15$\%,
which is comparable to the other estimates for $x=1.1-1.2$ ($m <$ 1
\msp) or $x=0.85$ ($m <$ 0.5 \msp) except for the IMF with $x$=1.6 for
the massive part in Table 2. This indicates that the IMF slope below 1
\ms is not much shallower than the slope above 1 \msp.

Our conclusion derived here is supported by the recent determination
of the IMF slope $x=1\pm 0.5$ for very low-mass stars ($m$ \ltsim 0.6
\msp) by M\'{e}ra et al. (1996) using both the updated mass-luminosity
relation by Chabrier, Baraffe, \& Plez (1996) and the luminosity
function where a contamination of unresolved binaries is carefully
taken into account (Kroupa 1995).

\acknowledgements

This work has been supported in part by the grant-in-Aid for
Scientific Research (05242102, 06233101, 07740194, 07804013) and
Center of Excellence (COE) research (07CE2002) of the Ministry of
Education, Science, and Culture in Japan.  F.M. acknowledges the
financial support from the Japan Society for the Promotion of Science
during her stay at the University of Tokyo.  We would like to thank
the referee, P. Kroupa, for many helpful comments that improved this
paper.


\begin{thebibliography}{}

\bibitem[]{}  
Arnett, W. D. 1978, ApJ, 219, 1008

\bibitem[]{}  
Bahcall, J. N. 1984, ApJ, 276, 169

\bibitem[]{}  
Bahcall, J. N. \& Soneira, R. M. 1980, ApJS, 44, 73

\bibitem[]{}  
Bahcall, J. N., Flynn, C., \& Gould, A. 1992, ApJ, 389, 234

\bibitem{} 
Barbuy, B. 1988, A\&A, 191, 121

\bibitem{} 
Barbuy, B., \& Erderlyi-Mendes, M. 1989, A\&A, 214, 239

\bibitem[]{}  
Carr, B. 1994, ARA\&A, 32, 531

\bibitem[]{}  
Chabrier, G., Baraffe, I., \& Plez, B. 1996, ApJ, 459, L91

\bibitem[]{}  
Charbonell, C., Meynet, G., Maeder, A., Schaller, G., \& Schaerer, D. 
1993, A\&AS, 101, 415

\bibitem{}
Edvardsson, B., Andersen, J., Gustafsson, B., Lambert, D. L.,
Nissen, P. E., \& Tomkin, J. 1993, A\&A, 275, 101

\bibitem[]{}  
Fazio, G. G., Dame, T. M., \& Kent, S. 1990, in IAU Symp. 139, 
The Galactic and Extragalactic Background Radiation, ed. S. Bower \& 
C. Leiner (Dordrecht:Kluwer), 35 

\bibitem[]{}  
Flynn, C., \& Fuchs, B. 1994, MNRAS, 270, 471

\bibitem[]{}  
Gould, A. 1990, MNRAS, 244, 25

\bibitem{}
Gratton, R. G. 1991, in IAU Symp. 145,  Evolution of Stars: The
Photospheric abundance Connection, ed.\ G. Michaud \& A. V. 
Tutukov (Montreal:Univ. Montreal), 27

\bibitem[]{}  
Hashimoto, M., Nomoto, K., \& Shigeyama, T. 1989, A\&A, 210, L5

\bibitem[]{}  
Hashimoto, M., Nomoto, K., Tsujimoto, T., \& Thielemann, F.-K. 1993a, in
Nuclei in the Cosmos, ed.\ K. K\"{a}ppeler \& K. Wisshak (Bristol:IOP
Publ.), 587

\bibitem[]{}  
Hashimoto, M., Iwamoto, K., \& Nomoto, K. 1993b, ApJ, 414, L105

\bibitem[]{}  
Hashimoto, M., Nomoto, K., Tsujimoto, T., \& Thielemann, F.-K. 1996, in
IAU Coll. 145, Supernovae and Supernova Remnants, ed.\ R. McCray \& Z.
Wang (Cambridge: Cambridge Univ. Press), 157

\bibitem[]{}   
Hill, R. J., Madore, B. F., \& Freedman, W. L. 1994, ApJ, 429, 204

\bibitem[]{}  
Hunter, D. A., Shaya, E. D., Holtzman, J. A., Light, R. M., O'Neil, E. J., 
\& Lynds, R. 1995, ApJ, 448, 179

\bibitem[]{}  
Kennicutt, R. C. 1983, ApJ, 272, 54

\bibitem[]{}  
Kennicutt, R. C., Tamblyn, P., \& Congdon, C. W. 1994, ApJ, 435, 22

\bibitem[]{}  
Kroupa, P. 1995, ApJ, 453, 350

\bibitem[]{}  
Kroupa, P., Tout, C. A., \& Gilmore, G. 1993, MNRAS, 262, 545

\bibitem[]{}  
Kuijken, K. 1995, in IAU Symp. 164, Stellar Populations, ed. P. C. 
van der Kruit \& G. Gilmore (Dordrecht:Kluwer), 195

\bibitem[]{}  
Kuijken, K., \& Gilmore, G. 1989, MNRAS, 239, 605

\bibitem[]{}  
Kuijken, K., \& Gilmore, G. 1991, ApJ, 367, L9

\bibitem[]{}  
Langer, N. 1989, A\&A, 220, 135

\bibitem[]{}  
Low, C., \& Lynden-Bell, D. 1976, MNRAS, 176, 367

\bibitem[]{}  
Maeder, A., 1990, A\&AS, 84, 139

\bibitem[]{}  
Maeder, A. 1992, A\&A, 264, 105

\bibitem[]{}  
Maeder, A. 1993, A\&A, 268, 833

\bibitem[]{}  
Massey, P., Lang, C. C., DeGioia-Eastwood, K., \& Garmany, C. D. 1995, 
ApJ, 438, 188

\bibitem[]{}  
M\'{e}ra, D., Chabrier, G., \& Baraffe, I. 1996, ApJ, 459, L87

\bibitem[]{}  
Miller, G. E., \& Scalo, J. M. 1979, ApJS, 41, 513

\bibitem{}
Nissen, P. E., \& Edvardsson, B. 1992, A\&A, 261, 255

\bibitem[]{}  
Nomoto, K. 1984, ApJ, 277, 791

\bibitem[]{}  
Nomoto, K. 1987, ApJ, 322, 206

\bibitem[]{}  
Nomoto, K., \& Hashimoto, M. 1988, Phys. Rep., 163, 13 

\bibitem[]{}  
Nomoto, K., Iwamoto, K., \& Suzuki, T. 1995, Phys. Rep., 256, 173

\bibitem[]{}  
Pagel, B. E. J., Simonson, E., Terlevich, R., \& Edmunds, M. 1992, 
MNRAS, 255, 325

\bibitem[]{}  
Price, N. M., \& Podsiadlowski, P., 1995, MNRAS, 273, 1041

\bibitem[]{}  
Renzini, A. \& Buzzoni, A. 1986, in The Spectral Evolution of Galaxies, 
ed. C. Chiosi \& A. Renzini (Dordrecht:Reidel), 195

\bibitem{}
Salpeter, E. E. 1955, ApJ, 121, 161

\bibitem[]{}  
Scalo, J. M. 1986, Fundam.Cosmic.Phys, 11, 1

\bibitem[]{}  
Schaerer, D., Meynet, G., Maeder, A., \& Schaller, G., 1993, A\&AS,
98, 523

\bibitem[]{}  
Silk, J. 1977, ApJ, 214, 152

\bibitem[]{}  
Silk, J. 1995, ApJ, 438, L41

\bibitem[]{}  
Sommer-Larsen, J. 1996, ApJ, 457, 118

\bibitem[]{}  
Thielemann, F.-K., Nomoto, K., \& Hashimoto, M. 1990, ApJ, 349, 222

\bibitem[]{}  
Thielemann, F.-K., Nomoto, K., \& Hashimoto, M. 1993, In Origin and
Evolution of the Elements, ed.\ N. Prantzos \& E. Vangoni-Flam
(Cambridge:Cambridge Univ. Press), 297

\bibitem[]{}  
Thielemann, F.-K., Nomoto, K., \& Hashimoto, M. 1994, in Supernovae,
Les Houches, Session LIV, ed.\ J. Audouze, S. Bludman, R.
Mochkovitch, \& J. Zinn-Justin (Amsterdam:Elsevier), 629

\bibitem[]{}  
Thielemann, F.-K., Nomoto, K., \& Hashimoto, M. 1996, ApJ, 460, 408

\bibitem[]{}  
Tinsley, B. M. 1980, Fund Cosmic Phys., 5, 287

\bibitem[]{}  
Tout, C. A., Pols, O. R., Eggleton, P. P., \& Han, Z. 1996, MNRAS, 281, 257

\bibitem[]{}
Tsujimoto, T., Nomoto, K., Yoshii, Y., Hashimoto, M., Yanagida, S.,
\& Thielemann, F.-K. 1995, MNRAS, 277, 945

\bibitem[]{}  
van den Heuvel, E. P. J. 1992, in Proc. European ISY'92 Conference

\bibitem[]{}  
van den Heuvel, E. P. J., \& Habets, G. M. H. J. 1984, Nature, 309, 698

\bibitem[]{}  
van den Kruit, P. C., 1986, A\&A, 157, 230

\bibitem[]{}  
Walker, T. P., Steigman, G., Schramm, D. N., Olive, K. A., \& Kang,
H.-S. 1991, ApJ, 376, 51

\bibitem[]{}  
Wang, B., Silk, J. 1993, ApJ, 406, 580

\bibitem[]{}  
Will, J.-M., Bomans, D. J., \& de Boer, K. S. 1995a, A\&A, 295, 54

\bibitem[]{} Will, J.-M., V\'{a}zquez, R. A., Feinstein, A., \&
Seggewiss, W. 1995b, A\&A, 301, 396

\bibitem[]{}  
Woosley, S. E., \& Weaver, T. A. 1982, in Essays in Nuclear 
Astrophysics, ed. C. A. Barnes, D. D. Clayton, \& D. N. Schramm 
(Cambridge:Cambridge Univ. Press), 377

\bibitem[]{} 
Woosley, S. E., \& Weaver, T. A. 1986, in IAU Colloq. 89, Radiation 
Hydrodynamics, ed. D. Mihalas \& K. H. Winkler (Dordrecht:Reidel), 91

\bibitem[]{}  
Woosley, S. E., Langer N., \& Weaver, T. A. 1993, ApJ, 411, 823

\bibitem[]{} 
Woosley, S. E., \& Weaver, T. A. 1995, ApJS, 101, 181

\bibitem{}   
Wyse, R. F. G., \& Gilmore, G. 1995, AJ, 110, 2771

\bibitem[]{}  
Yoshii, Y., \& Saio, H. 1985, ApJ, 295, 521

\bibitem[]{}  
Yoshii, Y., Ishida, K., \& Stobie, R. S. 1987, AJ, 93, 323

\bibitem[]{}  
Yoshii, Y., Tsujimoto, T., \& Nomoto, K. 1996, ApJ, 462, 266 (YTN)

\bibitem[]{}  
Young, J. S., \& Scoville, N. Z. 1991, ARA\&A, 91, 581

\end{thebibliography}
\end{document}